\documentclass[a4paper,twoside,12pt]{article}

\usepackage{graphicx,amstext,amsmath,amsbsy,amscd}
\usepackage{latexsym}
\usepackage{cite}
\usepackage{graphicx}

\newcommand{\bplus}{^{(+)}}
\newcommand{\bdplus}{_{(+)}}
\newcommand{\bminus}{^{(-)}}
\newcommand{\bdminus}{_{(-)}}

\newcommand{\gthree}{{\gamma_\ast}}

\newcommand{\cc}{^*}

\newcommand{\dbsp}{\displaybreak[0]\medsp}

\newcommand{\hatchi}{\hat{\chi}}
\newcommand{\hatbchi}{\Hat{\Bar{\chi}}}

\newcommand{\medsp}{\\[0.7ex]}

\newcommand{\tilchi}{\tilde{\chi}}
\newcommand{\tilbchi}{\Tilde{\Bar{\chi}}}

\newcommand{\uone}{\mbox{\slshape U}(1)}
\newcommand{\ve}{\varepsilon}

\newcommand{\dega}{\ensuremath{^\dag}}

\newcommand{\half}[1]{\ensuremath{\frac{#1}{2}}}

\newcommand{\inv}[1]{\ensuremath{\frac{1}{#1}}}

\newcommand{\Stext}[1]{\itindex{\mathcal{S}}{#1}}

\newcommand{\uhat}{\hat{u}}
\newcommand{\ubhat}{\Hat{\Bar{u}}}

\newcommand{\derfrac}[2][]{\frac{\partial #1}{\partial #2}}

\newcommand{\itindex}[2]{\ensuremath{#1_{\mbox{\scriptsize{\itshape #2}}}}}

\DeclareMathOperator{\extdm}{d}
\newcommand{\extd}{\extdm \!}

\begin{document}
\renewcommand{\thefootnote}{\fnsymbol{footnote}}
\thispagestyle{empty}
\begin{titlepage}

\begin{flushright}
  TUW-04-33\\
  hep-th/0411204\\
\end{flushright}
\vspace{1.21cm}

{\centering \textbf{\Large Two-Dimensional $\mathbf{N=(2,2)}$
    Dilaton Supergravity \\ from Graded Poisson-Sigma Models II:}\large
    \par}
\vspace{0.5cm}
{\centering \textbf{\large Analytic Solution and BPS States}\par}
\vspace{0.5cm}

\begin{center}
 L. Bergamin\footnote{bergamin@tph.tuwien.ac.at} and W. Kummer\footnote{wkummer@tph.tuwien.ac.at\par}
\end{center}

{\centering \textit{Institute for Theoretical Physics, Vienna University
of Technology}\par}

{\centering \textit{Wiedner Hauptstra{\ss}e 8-10, A-1040 Vienna, Austria}\par}
\vspace{0.5cm}
\begin{abstract}
The integrability of $N=(2,2)$ dilaton supergravity in two dimensions
is studied by the use of the graded Poisson Sigma model
approach. Though important differences compared to the purely bosonic
models are found, the general analytic solutions are obtained. The
latter include minimally gauged models as well as an ungauged version. BPS
solutions are an especially interesting subclass.
\end{abstract}

\end{titlepage}


\renewcommand{\thefootnote}{\arabic{footnote}}
\setcounter{footnote}{0}
\numberwithin{equation}{section}

\section{Introduction}
\label{sec:introduction}
Graded Poisson Sigma (gPSM) models \cite{Ikeda:1994dr,Izquierdo:1998hg,Strobl:1999zz,Ertl:2000si,Bergamin:2002ju,Bergamin:2003am} have proved to represent a powerful
formalism in the quest to solve several longstanding problems in 2d
(dilaton) supergravity \cite{Howe:1979ia,Park:1993sd}.
As shown in our first paper \cite{Bergamin:2004sr} on the subject
of $N=\left(2,2\right)$ supergravity \cite{Howe:1987ba,Alnowaiser:1990gh,Gates:1989ey,Gates:1989tn,Ketov:1994tb,Grisaru:1995dr,Grisaru:1995kn,Grisaru:1995dm,Gates:1996du,Nelson:1993vm,Gates:2000fj,Haack:2000di,Berkovits:2001tg} it is possible to formulate
the full actions including all fermionic contributions in a compact
form. Although our version works with a non-linear and open algebra, this turns
out to be no disadvantage thanks to the powerful symmetry principles of the
gPSM. In this way the very complicated and lengthy actions that follow from
superspace at the component level (cf.\ e.g.\
\cite{Grisaru:1995dr,Grisaru:1995kn,Grisaru:1995dm,Gates:1996du}) can be
avoided. It turns out that the existence of
the equivalent gPSM formulation is the key ingredient for
the numerous successes of this method which even extends to the quantization
of such theories \cite{Bergamin:2004us,Bergamin:2004aw}. On the other hand, there is an isomporphic
mapping of the symmetries as given by the superfield formulation for
$N=\left(2,2\right)$ onto the ones in the field content of gPSMs
\cite{Bergamin:2004sr}.

Our present paper is an immediate continuation of this work by using
another convenient feature of the gPSM formalism, namely the possibility
to derive the full classical solution, including the complete fermionic
contributions. Though many aspects of the calculation are a rather
straightforward generalization of previous results, new important problems
appear which are related to the question of the existence of Casimir-Darboux
coordinates on graded Poisson manifolds.

In order to provide a sufficiently self-contained presentation
we again start (Section \ref{sec:gPSM}) with a condensed description of the gPSM,
summarizing also the main results of \cite{Bergamin:2004sr} as needed for the application
to $N=\left(2,2\right)$ supergravity in the present work.
Section \ref{sec:solution} contains the solution for the chiral version of $N=\left(2,2\right)$
dilaton supergravity. It is enough to consider the case which corresponds
to vanishing kinetic term of the dilaton field in the version formulated
as a dilaton theory, because the general case can be obtained by straightforward
conformal transformation (Section \ref{sec:gensol}). The twisted chiral case is covered
as well by a simple {}``mirror-type'' redefinition of fields.
Section \ref{sec:ungauged} is devoted to a formulation of ungauged $N=\left(2,2\right)$
supergravity which consists in restricting the previous theory to
a fixed leaf of one of the Casimir functions in a gauge theory.
A short discussion of BPS solutions is the subject of Section \ref{sec:bps}, where
we show that (even in the so much simpler gPSM approach) the complications
for $N=\left(2,2\right)$ as compared to $N=\left(1,1\right)$ supergravity
\cite{Bergamin:2003mh} at present allow a consideration of the bosonic part only.
After the conclusion (Section \ref{sec:conclusions}) we decided to include as in \cite{Bergamin:2004sr}
again the Appendix describing the notation somewhat more in detail.

\section{gPSM for $\mathbf{N=(2,2)}$ supergravity}
\label{sec:gPSM}
In this Section some relevant formulae of (graded) Poisson Sigma
models and their application in dilaton supergravity are reviewed. For
further details ref.\ \cite{Bergamin:2004sr} and earlier literature on the
topic, esp.\
\cite{Schaller:1994es,Ertl:2000si,Grumiller:2002nm,Bergamin:2003am}
should be consulted.
A general gPSM consists of scalar fields
$X^I(x)$, which are themselves coordinates of a graded Poisson manifold with
Poisson tensor $P^{IJ}(X) = (-1)^{IJ+1} P^{JI}(X)$. The index
$I$, in the generic case, includes commuting as well as anti-commuting
fields\footnote{The usage of different indices as well as other features of
  our notation are explained in
  Appendix \ref{sec:notation}.
For further details one should consult ref.\ \cite{Ertl:2000si,Ertl:2001sj}.}. In addition one introduces the gauge
potential $A = \extd X^I A_I = \extd X^I A_{mI}(x) \extd x^m$, a one form with respect to the Poisson
structure as well as with respect to the 2d worldsheet coordinates. The gPSM
action reads\footnote{If the multiplication of forms is evident in what
  follows, the wedge symbol will be omitted.}
\begin{equation}
  \label{eq:gPSMaction}
  \begin{split}
    \Stext{gPSM} &= \int_M \extd X^I \wedge A_I + \half{1} P^{IJ} A_J \wedge
    A_I\ . 
  \end{split}  
\end{equation}
The Poisson tensor $P^{IJ}$ must have vanishing Nijenhuis tensor (obey a
Jacobi-type identity with respect to the Schouten bracket related as $\{ X^I,
X^J \} = P^{IJ}$ to the Poisson tensor)
\begin{equation}
\label{eq:nijenhuis}
 J^{IJK} =  P^{IL}\partial _{L}P^{JK}+ \mbox{\it 
g-perm}\left( IJK\right) = 0\ ,
\end{equation}
where the sum runs over the graded permutations. The variation of $A_I$ and $X^I$ in \eqref{eq:gPSMaction} yields the gPSM
equations of motion (eom-s)
\begin{align}
\label{eq:gPSMeom1}
  \extd X^I + P^{IJ} A_J &= 0\ ,\medsp
\label{eq:gPSMeom2}
  \extd A_I + \half{1} (\partial_I P^{JK}) A_K A_J &= 0\ .
\end{align}
Due to
\eqref{eq:nijenhuis} the action \eqref{eq:gPSMaction} is invariant under the
symmetry transformations
\begin{align}
\label{eq:symtrans}
  \delta X^{I} &= P^{IJ} \ve _{J}\ , & \delta A_{I} &= -\mbox{d} \ve
  _{I}-\left( \partial _{I}P^{JK}\right) \ve _{K}\, A_{J}\ ,
\end{align}
where the term $\extd \epsilon_I$ in the second of these equations provides
the justification for calling $A_I$ ``gauge fields''.


If the Poisson
tensor has a non-vanishing kernel
there exist (one or more) Casimir functions $C(X)$ obeying
\begin{equation}
\label{eq:casimir}
  \{ X^I, C \} = P^{IJ}\derfrac[C]{X^J} = 0\ ,
\end{equation}
which, when the $X^I$ obey the field equations of motion, are constants of
motion.

It was shown in ref.\ \cite{Bergamin:2004sr} that minimally gauged $N=(2,2)$ dilaton
supergravity can be described by a gPSM if the target space has four
(real) commuting dimensions, interpreted as a complex dilaton $X = \phi + i \pi$ and an auxiliary
field $X^a$, and four anti-commuting ones, which are combined in a complex two-component dilatino
$\chi^\alpha$. The associated gauge fields are the
spin-connection $\omega$, the
``zweibein'' $e_a$ and the complex gravitino $\psi_\alpha$. For a gauged
$\uone$ symmetry of $\chi^\alpha$ another $\uone$ gauge
field $B$ must be introduced. General
dilaton supergravity models are completely determined by two $X$-dependent
functions, namely a prepotential $u(X,\bar{X})$ and the conformal factor
$Q(X)$. The derivative of the latter is denoted as $Q'(X) = Z(X)$ and
defines the contributions quadratic in bosonic torsion. Furthermore it
is useful to introduce the potentials $w$ and $W$, which control the bosonic
and fermionic parts, respectively:
\begin{align}
  \label{eq:pcs7}
  w(X) &= \inv{4} e^{\bar{Q}/2} u & W(X,\bar{X}) &= - 2 w \bar{w}
\end{align}
With these definitions general
chiral dilaton supergravity is described by the Poisson tensor
\cite{Bergamin:2004sr}
\begin{gather}
\label{eq:psm10}
 \begin{alignat}{4}
   P^{a \phi} &= X^b {\epsilon_b}^a\ , &\qquad P^{\pi \phi} &= 0\ , &\qquad  P^{\alpha
   \phi} &= - \half{1} \chi^\beta \gthree{}_\beta{}^\alpha\ , &\qquad
   P^{\bar{\alpha} \phi} = - \half{1} \bar{\chi}^\beta
   \gthree{}_\beta{}^\alpha\ .
  \end{alignat}\medsp
\label{eq:psm11}
\begin{alignat}{3}
     P^{a \pi} &= 0 , &\qquad  P^{\alpha
     \pi} &= - \half{i} \chi^\beta \gthree{}_\beta{}^\alpha\ , &\qquad
     P^{\bar{\alpha} \pi} = \half{i} \bar{\chi}^\beta
     \gthree{}_\beta{}^\alpha\ .
\end{alignat}\medsp
  \label{eq:pcs8.1}
  P^{ab} = \epsilon^{ab} \Bigl( e^{-(Q + \bar{Q})/2} W' + \half{1} Y (Z +
  \bar{Z}) + \inv{4} \chi^2 e^{-Q/2} \bar{w}'' + \inv{4}\bar{\chi}^2
  e^{-\bar{Q}/2} w'' \Bigr)\ , \medsp
\label{eq:pcs8.2}
  P^{a \alpha} = i e^{-\bar{Q}/2} w'
  (\bar{\chi}\gamma^a)^\alpha - \frac{\bar{Z}}{4} X^b (\chi \gamma_b \gamma^a
  \gthree)^\alpha\ , \medsp 
\label{eq:pcs8.3}
 P^{a \bar{\alpha}} = i e^{-Q/2} \bar{w}'
  (\chi\gamma^a)^\alpha - \frac{Z}{4} X^b (\bar{\chi} \gamma_b \gamma^a
  \gthree)^\alpha\ , \medsp
\label{eq:pcs8.4}
  P^{\alpha \bar{\beta}} = - 2 i X^a (\gamma_a)^{\alpha \beta}\ , \medsp
\label{eq:pcs8.5}
  \begin{align} P^{\alpha \beta} &= (u + \frac{\bar{Z}}{4} \chi^2)
  \gthree^{\alpha \beta}\ , &  P^{\bar{\alpha} \bar{\beta}} &= (\bar{u} + \frac{Z}{4} \bar{\chi}^2)
  \gthree^{\alpha \beta}\ . \end{align}
\end{gather}
The bosonic part of the Poisson tensor has four dimensions but at most
rank two. Therefore there exist at least two (real) commuting Casimir
functions, which can be chosen as
\begin{gather}
\label{eq:pcs9.1}
  C = 8 \bigl(W +e^{(Q + \bar{Q})/2} ( Y + \inv{4} \chi^2 e^{-Q/2} \bar{w}' +
  \inv{4} \bar{\chi}^2 e^{-\bar{Q}/2} w')\bigr)\ , \medsp
\label{eq:pcs9.2}
  C_\pi = \pi + i e^{\bar{Q}/2} \frac{\bar{w}}{C} \chi^2 - e^{Q/2}\frac{w}{C}
  \bar{\chi}^2 - \frac{e^{(Q+\bar{Q})/2}}{C} X^a (\chi \gamma_a \gthree
  \bar{\chi})\ . 
\end{gather}
The first one is related to the energy\footnote{This energy conservation is a
  pecular feature of 2d (super-)gravity, even in the presence of matter \cite{Kummer:1995qv,Grumiller:1999rz,Bergamin:2003mh}.} the second to the $\uone$
charge of the system.

An important simplified model is dilaton prepotential supergravity
\cite{Ertl:2000si} obtained for the special case
$Z=0$ (cf.\ Section 3 of \cite{Bergamin:2004sr}). General supergravity models can be obtained from the latter by
means of conformal transformations, which are interpreted as
target-space diffeomorphisms. Therefore, for any \emph{local} analysis
it is sufficient as a first step to consider this simpler class of
models. Nevertheless, the conformal transformations towards $Z\neq0$ need not be
defined globally and thus the latter models tend to be physically
inequivalent.

Finally we note that under the exchange $\chi^- \leftrightarrow
\bar{\chi}^-$ and $\psi_- \leftrightarrow \bar{\psi}_-$ a chiral
gauging of the internal $\uone$ turns into a twisted chiral one. This map represents mirror symmetry; it is defined globally and thus
physics do not change, as is expected for the
geometric part of the action.

\section{Solution of dilaton prepotential SUGRA}
\label{sec:solution}
The aim of this work is to study the integrability of $N=(2,2)$
dilaton supergravity and to derive its analytic solution. As announced above
the explicit
calculations can be restricted to the chiral version of dilaton
prepotential supergravity, the general theories are then obtained by
means of conformal transformations. Although we
find agreement with the general statements about (g)PSMs that the
models developed in this work are integrable, important differences
between graded PSMs and ordinary (bosonic) PSMs become manifest here.
\subsection{gPSM and Casimir-Darboux coordinates}
\label{sec:solution.1}
The integrability of bosonic dilaton gravity may be checked by
explicit integration of the equations of motion \cite{Kummer:1995qv}. However, once the
theory is formulated in terms of a PSM this characteristic is
guaranteed by the fact that any Poisson manifold locally can be
transformed to Casimir-Darboux (CD) coordinates\footnote{Such coordinates exist on regular sheets of the Poisson
  manifold, only. Solutions on irregular sheets have to be considered
  seperately \cite{Klosch:1996fi,Strobl:1999wv}. In the bosonic model these are restricted
  to the point $X^{++}=X^{--}=0$. Such solutions describe constant
  dilaton vacua or a bifurcation point. Some solutions of this type are
  discussed in Section \ref{sec:bps}.}. As the integrability of the model at hand is not
obvious in their ``physical'' coordinates, it is helpful to choose new
coordinates that are CD or at least almost CD.

In the following we assume that $X^{++} \neq 0$. Then for the purely bosonic
 theory one can choose
 the  Casimir-Darboux coordinates $\{C,
\pi, \phi, \lambda\}$ with $\lambda = - \ln |X^{++}|$. The only
non-vanishing bracket among these variables is $\{\lambda, \phi\}=
1$. As $\pi$ does not commute with the fermions, this choice does not
lead to CD coordinates for the full theory, but they would be  $\{C,
C_\pi, \phi, \lambda\}$ plus some convenient choice for the
fermions. However the former choice turns out to be the preferrable
one: First, the replacement $\pi \rightarrow C_\pi$ leads to lengthy
 equations (cf.\ Section 7 of \cite{Bergamin:2004sr})  and second, solutions for $C_B=0$
 cannot be obtained in this way, as $C_\pi$ contains inverse powers in
 this function.

Among the fermionic coordinates we follow the idea of ref.\ \cite{Ertl:2000si}
to choose the Lorentz invariant quantities\footnote{Throughout Section
  \ref{sec:solution} variables like $\chi^\pm$ etc.\  refer to the restricted
  case $Z=0$. In the transition to $Z\neq0$ in Section \ref{sec:gensol} we
  shall rename the variables of Section \ref{sec:gPSM} by underlining them.}
\begin{align}
\label{eq:sol1.1}
  \tilchi\bplus &= \inv{\sqrt{|X^{++}|}} \chi^+\ , & \tilbchi\bplus &= \inv{\sqrt{|X^{++}|}} \bar{\chi}^+\ , \dbsp
\label{eq:sol1.2}
  \hatchi\bminus &= \sqrt{|X^{++}|} \chi^- - \frac{i \sigma u}{2 \sqrt{2}}
  \tilbchi\bplus\ , & \hatbchi\bminus &= \sqrt{|X^{++}|} \bar{\chi}^- -
  \frac{i \sigma \bar{u}}{2 \sqrt{2}} \tilchi\bplus\ ,
\end{align}
as new coordinates ($\sigma$ denotes the sign\footnote{According to the
 conventions outlined in the Appendix, quantities like $X^{\pm\pm}$ are
 imaginary (cf.\ eqs.\ \eqref{eq:A10} and \eqref{eq:gammalc}). In
 refs.\ \cite{Ertl:2000si,Bergamin:2003am} a real value for $X^{++}$ has been
 assumed, which corresponded to a slightly different convention.} of $i X^{++}$). The second term in the definition of $\hatchi\bminus$ is motivated by the bracket
\begin{equation}
  \label{eq:sol2}
  \{ \lambda, \sqrt{|X^{++}|} \chi^-\} =  \{ \lambda, \tilchi\bminus \} =
  \frac{i \sigma}{2 \sqrt{2}} u' \tilbchi\bplus\ .
\end{equation}
It is now straightforward to check that the Poisson brackets---beside the purely bosonic ones already mentioned above---reduce to
\begin{align}
\label{eq:sol3.11}
  \{\pi, \tilchi\bplus \} &= \half{i} \tilchi\bplus\ , & \{\pi, \tilbchi\bplus
  \} &=  - \half{i} \tilbchi\bplus\ ,\dbsp
\label{eq:sol3.12}
  \{\pi, \hatchi\bminus \} &= - \half{i} \hatchi\bminus\ , & \{\pi,
  \hatbchi\bminus \} &=  \half{i} \hatbchi\bminus\ , \dbsp
\label{eq:sol3.2}
 \{ \tilchi\bplus, \tilbchi\bplus\} &= - 2 \sqrt{2} i \sigma\ , & \{
 \hatchi\bminus, \hatbchi\bminus\} &= -\frac{i \sigma}{2 \sqrt{2}} C\ , 
\end{align}
while all remaining brackets are zero. All details of the model are
hidden in the redefinition of the fields, and the equations of motion
for the new variables become independent of the prepotential $u$. To
distinguish the set of transformed gauge potentials from the original
ones they all are denoted with a tilde $(\tilde{A}_C, \tilde{A}_\pi,
\tilde{A}_\lambda, \tilde{A}_\phi, \tilde{A}\bdplus,
\Tilde{\Bar{A}}\bdplus, \tilde{A}\bdminus,
\Tilde{\Bar{A}}\bdminus)$. Also, the action \eqref{eq:gPSMaction} is
expressed in terms of the transformed Poisson tensor related to the
brackets \eqref{eq:sol3.11}-\eqref{eq:sol3.2} and $\{\lambda,\phi\}=1$.
Variation of the action with respect to these $\tilde{A}_I$ yields the eom-s
\begin{gather}
\label{eq:1A}
\extd C = 0\ ,\medsp
\label{eq:1B}
  \extd \pi + \half{i} (\tilchi\bplus \tilde{A}\bdplus - \tilbchi\bplus
  \Tilde{\Bar{A}}\bdplus - \hatchi\bminus \tilde{A}\bdminus + \hatbchi\bminus
  \Tilde{\Bar{A}}\bdminus ) = 0\ ,\medsp
\label{eq:1C}
\begin{alignat}{2}
  \extd \phi - \tilde{A}_\lambda &= 0\ , &\qquad
  \extd \lambda + \tilde{A}_\phi &= 0\ ,
\end{alignat} \medsp
\label{eq:1E}
\begin{alignat}{2}
\extd \tilchi\bplus + 2 \sqrt{2} i \sigma \Tilde{\Bar{A}}\bdplus - \half{i}
\tilchi\bplus \tilde{A}_\pi &= 0\ , &\qquad
\extd \tilbchi\bplus + 2 \sqrt{2} i \sigma \tilde{A}\bdplus + \half{i}
\tilbchi\bplus \tilde{A}_\pi &= 0\ ,
\end{alignat}
\medsp
\label{eq:1G}
\begin{alignat}{2}
\extd \hatchi\bminus - \frac{i \sigma}{2 \sqrt{2}} C \Tilde{\Bar{A}}\bdminus +
\half{i} \hatchi\bminus \tilde{A}_\pi &= 0\ , &\qquad
\extd \hatbchi\bminus - \frac{i \sigma}{2 \sqrt{2}} C \tilde{A}\bdminus -
\half{i} \hatbchi\bminus \tilde{A}_\pi &= 0\ ,
\end{alignat}
\end{gather}
while variation with respect to $X^I$ produces
\begin{gather}
  \label{eq:2A}
\extd \tilde{A}_C - \frac{i \sigma}{2 \sqrt{2}} \tilde{A}\bdminus \wedge
\Tilde{\Bar{A}}\bdminus = 0\ , \medsp
\label{eq:2B}
\begin{alignat}{3}
\extd \tilde{A}_\pi &= 0\ , &\qquad
\extd \tilde{A}_\phi &= 0 \, &\qquad
\extd \tilde{A}_\lambda &= 0\ ,
\end{alignat}
\medsp
\label{eq:2E}
\begin{alignat}{2}
\extd \tilde{A}\bdplus + \half{i} \tilde{A}\bdplus\wedge \tilde{A}_\pi &= 0\ , &\qquad
\extd \Tilde{\Bar{A}}\bdplus - \half{i} \Tilde{\Bar{A}}\bdplus\wedge
\tilde{A}_\pi &= 0\ ,
\end{alignat}
\medsp
\label{eq:2G}
\begin{alignat}{2}
\extd \tilde{A}\bdminus - \half{i} \tilde{A}\bdminus\wedge \tilde{A}_\pi &= 0\
, &\qquad
\extd \Tilde{\Bar{A}}\bdminus + \half{i} \Tilde{\Bar{A}}\bdminus\wedge
\tilde{A}_\pi &= 0\ .
\end{alignat}
\end{gather}
It is obvious from the definition of $C_\pi$ in \eqref{eq:pcs9.2} and
the second equation in \eqref{eq:sol3.2} that complications arise if the
body of the Casimir function $C$ vanishes. A similar problem already appears
in $N=(1,1)$ supergravity \cite{Ertl:2000si} and---as a concise discussion of
this point has
not been given so far---this simpler model is considered first.
As is
easily seen from the above result by reducing configuration space to
the one of $N=(1,1)$, CD coordinates are obtained after
simple rescalings by negative powers of $\sqrt{C}$ except for the bracket in \eqref{eq:sol3.2}
(cf.\ Section 8 in \cite{Ertl:2000si}, unimportant constants and
factors of $i$ are omitted)
\begin{equation}
  \label{eq:soln1}
  \{ \hat{\chi}\bminus, \hat{\chi}\bminus \} = C 
\end{equation}
and the respective eom-s
\begin{align}
  \label{eq:soln2}
  \extd \hat{\chi}\bminus - C \tilde{A}\bdminus &= 0\ , & \extd
  \tilde{A}\bdminus &=0\ .
\end{align}
Clearly \eqref{eq:soln2} can be integrated for any value of
$C = \mbox{const.}$ $\tilde{A}\bdminus$ is closed and thus locally
$\tilde{A}\bdminus = \extd \zeta\bdminus$. For constant $C$
the simple equation $ \hat{\chi}\bminus = C \zeta\bdminus + \hat{\chi}\bminus_0$
is obtained with arbitrary $\zeta\bdminus$ and a constant $\hat{\chi}\bminus_0$. Nevertheless one has to distinguish three different
cases:
\begin{enumerate}
\item $C=0$. In that case the first equation in \eqref{eq:soln2} defines
  $\hat{\chi}\bminus$ as a second anti-commuting Casimir function $\hat{\chi}\bminus_0$
  \cite{Ertl:2000si}, $\tilde{A}\bdminus$ is the associated gauge
  potential.
\item $C\neq0$. This case is divided into two sub-classes:
  \begin{enumerate}
  \item Non-vanishing body of $C$. Then after a rescaling of
  $\hat{\chi}^-$ by $1/\sqrt{C}$ in \eqref{eq:soln1} CD coordinates are obtained with $C$
  being the only Casimir function. Obviously the solution for
  $\tilde{A}\bdminus$ can be expressed in terms of
  $\hat{\chi}\bminus$.
  \item Vanishing body of $C$. This is a subtle case having no
  counterpart in the purely bosonic model. On the one hand,
  $\hat{\chi}\bminus$ is not a Casimir function as for $C=0$, but on the other hand
  a division by $C$ is excluded as $C^{-1}$ does not exist. Therefore
  it is impossible to transform the system to CD coordinates
  and to express the solution for $\tilde{A}\bdminus$ in terms of the
  target-space coordinates. From the first equation in \eqref{eq:sol2}
  it even seems that the solution for $\hat{\chi}\bminus$ depends on
  $\zeta\bdminus$. For the special case at hand it can be argued on
  general grounds that this is not the case \cite{Bergamin:2003mh}: As
  the solutions are parametrized by only two anti-commuting variables, $C
  \tilde{A}\bdminus$ must vanish if $C$ is pure soul. It is important
  to notice that this is a fortunate accident in this case.
  \end{enumerate}
\end{enumerate}
\subsection{Integration for generic Casimir function}
\label{sec:solution.2}
We now show that a general solution can be obtained with the only
restriction $X^{++} \neq 0$ (or equivalently $X^{--}\neq 0$.)
The solutions of \eqref{eq:1C} and \eqref{eq:2B} are immediate with
$\tilde{A}_\lambda = \extd \phi$, $\tilde{A}_\phi = - \extd \lambda$ and
$\tilde{A}_\pi= - \extd F_\pi$, where $F_\pi$ is a free function. If $C$ has
non-vanishing body we could solve all four equations in
\eqref{eq:1E}-\eqref{eq:1G} for the fermionic gauge potentials. Nevertheless
we want to proceed in a different way, because solutions with $C=0$ or $C=\mbox{pure soul}$
do appear in this model and in particular represent the interesting class of
BPS states (cf.\ \cite{Bergamin:2003mh} and Section \ref{sec:bps} below). From \eqref{eq:1E} we obtain
\begin{align}
  \label{eq:sol5}
  \tilde{A}\bdplus &= \frac{i \sigma}{2\sqrt{2}} \extd \tilbchi\bplus +
  \frac{\sigma}{4 \sqrt{2}} \tilbchi\bplus \extd F_\pi\ , &
  \Tilde{\Bar{A}}\bdplus &= \frac{i \sigma}{2\sqrt{2}} \extd
  \tilchi\bplus - \frac{\sigma}{4 \sqrt{2}} \tilchi\bplus \extd
  F_\pi\ .
\end{align}
For the remaining fermionic variables the equations in \eqref{eq:2G} must be addressed first. They imply
\begin{align}
  \label{eq:sol6}
  \tilde{A}\bdminus &= \extd \zeta\bdminus + \half{i} \zeta \bdminus \extd
  F_\pi\ , &  \Tilde{\Bar{A}}\bdminus &= \extd \bar{\zeta}\bdminus - \half{i} \bar{\zeta} \bdminus \extd F_\pi
\end{align}
for some complex anti-commuting function $\zeta\bdminus$. Now these solutions are
inserted into the equations in \eqref{eq:1G} that yield after integration
\begin{align}
\label{eq:sol7.1}
  \hatchi\bminus &= \frac{i \sigma}{2\sqrt{2}} C \bar{\zeta}\bdminus  +
  e^{\half{i} F_\pi} \lambda_0\bminus\ , & \hatbchi\bminus &= \frac{i \sigma}{2\sqrt{2}} C \zeta\bdminus  + e^{-\half{i}
  F_\pi} \bar{\lambda}_0\bminus\ . 
\end{align}
In this solution $\lambda_0\bminus$ is a constant spinor. Among the variations
with respect to $\tilde{A}_I$ there remains \eqref{eq:1B}, which should
produce by  $\extd C_\pi = 0$
the constant of motion. When \eqref{eq:sol5} and \eqref{eq:sol7.1} are
inserted into that equation indeed a total derivative is obtained. Its
integration with some integration constant $C_\pi^0$
yields\footnote{\label{foot:1}Anticipating the result of Section \ref{sec:sol.1} the
  constant $C_\pi^0$ cannot be set to zero in order to match the prescription
  in \eqref{eq:pcs9.2}.}
\begin{equation}
  \label{eq:sol8}
  C_\pi = \pi - \frac{\sigma}{4 \sqrt{2}} \tilchi\bplus \tilbchi\bplus -
  \frac{\sigma}{4 \sqrt{2}} C \zeta\bdminus \bar{\zeta}\bdminus - \half{i}
  (e^{\half{i} F_\pi} \lambda_0\bminus \zeta\bdminus - e^{-\half{i} F_\pi}
  \bar{\lambda}_0\bminus \bar{\zeta}\bdminus ) + C_\pi^0\ .
\end{equation}
It remains to find the explicit form of $\tilde{A}_C$ from \eqref{eq:2A} and \eqref{eq:sol6}. This gauge potential depends on an additional free function $\extd F$ and after a straightforward integration can be written as
\begin{equation}
  \label{eq:sol9}
  \tilde{A}_C = - \extd F + \frac{i \sigma}{4 \sqrt{2}} \bigl( i \zeta\bdminus
  \bar{\zeta}\bdminus \extd F_\pi - (\zeta\bdminus \extd \bar{\zeta}\bdminus +
  \bar{\zeta}\bdminus \extd \zeta\bdminus) \bigl)\ .
\end{equation}

Before proceeding to the discussion of specific classes of solutions we should
worry about the transformations back to the original ``physical''
coordinates. The one of the gauge potentials follows straightforwardly by applying target space diffeomorphisms: $A_I = \partial
\tilde{X}^J / \partial X^I \tilde{A}_J$. The $\tilde{X}^I$ comprise the CD
coordinates of the bosonic sector as defined in the second paragraph of
Section \ref{sec:solution.1} together with the fermionic components
\eqref{eq:sol1.1} and \eqref{eq:sol1.2}. The explicit result reads:
\begin{gather}
  \label{eq:sol10.1}
  \omega = \frac{\extd X^{++}}{X^{++}} + \bigl( - (\bar{u} u)' + \chi^- \chi^+
  \bar{u}'' + \bar{\chi}^- \bar{\chi}^+ u''\bigr) \tilde{A}_C - \frac{i \sigma
  u'}{2 \sqrt{2}} \tilbchi\bplus \tilde{A}\bdminus - \frac{i \sigma \bar{u}'}{2 \sqrt{2}} \tilchi\bplus \Tilde{\Bar{A}}\bdminus\medsp
\label{eq:sol10.2}
  B = - \extd F_\pi - i \bigl( (u' \bar{u} - u \bar{u}') + \chi^- \chi^+
  \bar{u}'' - \bar{\chi}^- \bar{\chi}^+ u''\bigr) \tilde{A}_C  + \frac{\sigma
  u'}{2 \sqrt{2}} \tilbchi\bplus \tilde{A}\bdminus - \frac{\sigma \bar{u}'}{2 \sqrt{2}} \tilchi\bplus \Tilde{\Bar{A}}\bdminus \medsp
  \label{eq:sol10.3}
  \begin{split}
    e_{++} &= - \frac{\extd \phi}{X^{++}} + 8 X^{--} \tilde{A}_C  - \inv{2 X^{++}} \bigl( \tilchi\bplus \tilde{A}\bdplus + \tilbchi\bplus \Tilde{\Bar{A}}\bdplus\medsp
 &\quad - (\tilchi\bminus + \frac{i \sigma u}{2 \sqrt{2}} \tilbchi\bplus)
 \tilde{A}\bdminus -  (\tilbchi\bminus + \frac{i\sigma \bar{u}}{2 \sqrt{2}} \tilchi\bplus) \Tilde{\Bar{A}}\bdminus \bigr)
  \end{split}\medsp
\label{eq:sol10.4}
e_{--} = 8 X^{++} \tilde{A}_C \medsp
\label{eq:sol10.5}
\psi_+ = \inv{\sqrt{|X^{++}|}} \bigl(\tilde{A}\bdplus - \frac{i\sigma \bar{u}}{2 \sqrt{2}} \Tilde{\Bar{A}}\bdminus \bigr) - \bar{u}' \chi^- \tilde{A}_C \medsp
\label{eq:sol10.6}
\bar{\psi}_+ = \inv{\sqrt{|X^{++}|}} \bigl(\Tilde{\Bar{A}}\bdplus - \frac{i\sigma u}{2 \sqrt{2}} \tilde{A}\bdminus \bigr) - u' \bar{\chi}^- \tilde{A}_C \medsp
\label{eq:sol10.7}
\psi_- = \sqrt{|X^{++}|} \tilde{A}\bdminus + \bar{u}' \chi^+ \tilde{A}_C \medsp
\label{eq:sol10.8}
\bar{\psi}_- = \sqrt{|X^{++}|} \Tilde{\Bar{A}}\bdminus + u' \bar{\chi}^+ \tilde{A}_C
\end{gather}
The dependence on a specific model is determined by the prepotential $u(\phi)$
alone.

The similarity of this solution to the one of bosonic gravity
\cite{Grumiller:2002nm} as well as of $N=(1,1)$ supergravity
\cite{Ertl:2000si,Bergamin:2003am} is immediate. In the following we want to
discuss more in detail the solution of dilaton prepotential supergravity for
different values of the Casimir function $C$. The solution of the general
supergravity model \eqref{eq:pcs7}-\eqref{eq:pcs9.2} with $Z\neq0$ then simply
follows by applying certain target space diffeomorphisms onto these solutions.

\subsection{Non-vanishing body of the Casimir $\mathbf{C}$}
\label{sec:sol.1}

The simplest solution is obtained on a patch with non-vanishing body of the
Casimir function, as in that case $C^{-1}$ is well defined. Then also the
definition of the second Casimir \eqref{eq:pcs9.2}, taken at $Q=0$, makes sense.

Thus we can solve still the equations in \eqref{eq:sol7.1} for $\bar{\zeta}\bdminus$ and $\zeta\bdminus$, resp.:
\begin{align}
  \label{eq:sol11}
  \zeta\bdminus &= - \frac{2 \sqrt{2} i \sigma}{C} (\hatbchi\bminus -
  e^{-\half{i} F_\pi} \bar{\lambda}_0\bminus) & \bar{\zeta}\bdminus &= -
  \frac{2 \sqrt{2} i \sigma}{C} (\hatchi\bminus - e^{\half{i} F_\pi} \lambda_0\bminus) 
\end{align}
It is seen that the $\lambda_0\bminus$ part drops out of the solution for $\tilde{A}\bdminus$:
\begin{align}
  \label{eq:sol12}
  \tilde{A}\bdminus &= - \frac{2 \sqrt{2} i \sigma}{C} (\extd \hatbchi\bminus +
  \half{i} \hatbchi\bminus \extd F_\pi) &  \Tilde{\Bar{A}}\bdminus &= -
  \frac{2 \sqrt{2} i \sigma}{C} (\extd \hatchi\bminus - \half{i} \hatchi\bminus \extd F_\pi)
\end{align}
Of course, one could proceed by integrating again \eqref{eq:1B} and \eqref{eq:2A} with these formulas. Instead we insert \eqref{eq:sol11} into the expressions \eqref{eq:sol8} and \eqref{eq:sol9} which leads to
\begin{gather}
  \label{eq:sol13}
  C_\pi = \pi - \frac{\sigma}{4 \sqrt{2}} \tilchi\bplus \tilbchi\bplus -
  \frac{\sqrt{2} \sigma}{C}(\hatchi\bminus \hatbchi\bminus -
  \lambda_0\bminus \bar{\lambda}_0\bminus)\ , \medsp
\label{eq:sol14}
\begin{split}
  \tilde{A}_C &= -\extd\bigl(F + \frac{\sqrt{2} i \sigma}{C^2} (e^{\half{i} F_\pi} \lambda_0\bminus \hatbchi\bminus +e^{-\half{i} F_\pi} \bar{\lambda}_0\bminus \hatchi\bminus ) \bigr)\medsp
  &\quad - \frac{\sqrt{2} \sigma}{C^2} \bigr(\hatchi\bminus \hatbchi\bminus
  \extd F_\pi +i (\hatbchi\bminus \extd \hatchi\bminus +\hatchi\bminus \extd
  \hatbchi\bminus) \bigl)\ .
\end{split}
\end{gather}
Starting instead with the definitions \eqref{eq:sol12} one obtains the
same result up to the terms dependent on $\lambda_0\bminus$, which are
clearly absent in that case. However, in \eqref{eq:sol13} the last term simply
produces the constant $C_\pi^0$ in \eqref{eq:sol8}. In \eqref{eq:sol14} the
terms $\propto\lambda_0\bminus$ can be absorbed by a
redefinition of $F$.

To summarize: this solution is parametrized by the two Casimir
functions according to \eqref{eq:pcs9.1} and \eqref{eq:pcs9.2}, by the
associated ``gauge potentials'' $\extd F$ and $\extd F_\pi$ as well as
by the target space variables $\phi$, $\lambda$, $\tilchi\bplus$,
$\tilbchi\bplus$, $\hatchi\bminus$ and $\hatbchi\bminus$. In the
general solution the
spinorial gauge potentials are determined by \eqref{eq:sol5} and
\eqref{eq:sol12}, $\tilde{A}_C$ is given by \eqref{eq:sol14} and $\pi$
inside the prepotential must be expressed by $C_\pi$ and spinorial
terms according to \eqref{eq:sol13}.
\subsection{Vanishing Casimir $\mathbf{C}$}
\label{sec:sol.2}
The other extreme is the case $C\equiv0$. Then the definition of
$\hatchi\bminus$ decouples completely from $\bar{\zeta}\bdminus$. This
implies that the latter spinors cannot be expressed in terms of the
target space variables, the typical situation one encounters if the
target space coordinate is a Casimir function of the Poisson manifold. Indeed,
as shown in Section \eqref{sec:solution.1} a fermionic Casimir function occurs
for $C=0$ in the $N=(1,1)$ case \cite{Ertl:2000si}, which may simply be
identified with $\hatchi\bminus$. At least in that limit this result should be
reproduced here. Nevertheless it is obvious from \eqref{eq:sol7.1} that $\extd
\hatbchi\bminus \neq 0$ irrespective of the value of $C$. The new constant of
motion coincides with the (for $C\neq0$ irrelevant) constant $\lambda\bminus_0$:
\begin{align}
  \extd \lambda_0\bminus &= \extd(e^{-\half{i} F_\pi} \hatchi\bminus)
  = 0 &
  \extd \bar{\lambda}_0\bminus &= \extd(e^{\half{i} F_\pi} \hatbchi\bminus) =
  0
\end{align}
On the one hand, this result has the expected property to reduce to the one
found in $N=(1,1)$ in the limit where the target space is reduced to this
theory. On the other hand, the constant of motion cannot be expressed
completely in terms of the target space variables because $F_\pi$ is related
to a gauge field.
This pecularity appears in the definition of the ``second Casimir'' $C_\pi$
as well. Indeed for $C=0$ the definition \eqref{eq:pcs9.2} is ill defined as
there appear inverse powers of $C$. Of course the combination $C \cdot C_\pi$
is a well defined Casimir, but in the limit $C\rightarrow0$ it makes no sense
as $C$ and $C \cdot C_\pi$ are no longer independent. The correct solution is
found by looking at the equations of motion including the gauge fields:
Indeed they could be  integrated in full generality in eq.\  \eqref{eq:sol8}; for $C=0$ we find
\begin{equation}
  \label{eq:sol15}
   C_\pi = \pi - \frac{\sigma}{4 \sqrt{2}} \tilchi\bplus
   \tilbchi\bplus - \half{i} (e^{\half{i} F_\pi} \lambda_0\bminus
   \zeta\bdminus - e^{-\half{i} F_\pi} \bar{\lambda}_0\bminus
   \bar{\zeta}\bdminus )\ .
\end{equation}
It should be noted that the last term depends on $F_\pi$ and $\zeta\bdminus$, i.e.\ quantities that are not part of the target space.
All remaining gauge potentials follow straightforwardly from the result
obtained already above.

Obviously, all problems of finding CD coordinates for $C=0$ are intimately
connected with the divergences at $C=0$ that show up in $C_\pi$.
As we work throughout with an explicit basis on the Poisson manifold we should
ask whether the characteristics of our solution are generic or a pecularity of
our choice of coordinates. By analyzing this the meaning of the
(non-)existence of CD coordinates should become more transparent.

The first question concerns the existence of a Casimir function. Indeed, a
very simple solution for the elimination of the divergences at $C\rightarrow0$
 in \eqref{eq:pcs9.2} could
be that then a second (commuting) Casimir exists for vanishing fermions
only. However, the analysis of this Section showed that there exist for all
solutions at least two commuting constants of motion, one related to $C$ the
other one related to $C_\pi$. Therefore it remains to check, whether new
coordinates can be chosen in such a way that the Casimir function $C_\pi$
remains regular. The problem can be considered in two different versions:
\begin{enumerate}
\item One can ask whether such coordinates exist in an entire neighborhood of
  $C=0$. As such a region includes points where $C\neq0$ has non-vanishing
  body, one can reduce this to the question whether there exist
  coordinates
  such that $C_\pi$ remains regular in the limit $C=0$. Indeed, such
  coordinates can be defined for a restricted range of the fermionic
  fields. This is most easily seen from equation \eqref{eq:sol13},
  where by the rescaling
  \begin{align}
    \label{eq:soln10}
    \hatchi\bminus &= \sqrt{C} \check{\chi}\bminus\ , &  \hatbchi\bminus &= \sqrt{C} \Check{\Bar{\chi}}\bminus
  \end{align}
  all divergences from $C_\pi$ disappear.
  However from the definition of $\hatchi$ in eq.\ \eqref{eq:sol1.2} it is
  seen that this implies in the limit of $C\rightarrow0$
  \begin{align}
    \label{eq:soln11}
    \chi^- &= \inv{2 \sqrt{2}} \frac{u}{X^{++}} \bar{\chi}^+\ ,  &
    \bar{\chi}^- &= \inv{2 \sqrt{2}} \frac{\bar{u}}{X^{++}} \chi^+\ .
  \end{align}
  Clearly, the general solution from \eqref{eq:sol7.1} with independen
  $\chi\bminus$ and $\chi\bplus$ need not respect this
  constraint for $C=0$.
\item A weaker requirement would be to find regular coordinates that are valid on the
  sheet $C=0$ only. Here a similar problem as in the example of Section
  \ref{sec:solution.1} is encountered. Clearly, the system in eqs.\
  \eqref{eq:sol3.11}-\eqref{eq:sol3.2} cannot be transformed to CD coordinates
  as the inverse of a spinorial quantity is not defined (remember that the
  remaining coordinates are already CD).
\end{enumerate}

There is yet another way to illustrate the difference with respect to the solutions
with $C\neq0$: The solution with $C=0$ is parametrized by $C=0$, $\extd
F$, $C_\pi$, $F_\pi$, $\phi$, $\lambda$, $\tilde{\chi}\bplus$,
$\Tilde{\bar{\chi}}\bplus$, $\lambda_0\bminus$,
$\bar{\lambda}_0\bminus$, $\zeta\bdminus$,
$\bar{\zeta}\bdminus$. Counting the degrees of freedom it is seen that
this configuration space is by one real bosonic and one complex fermionic constant larger than
the one found for $C\neq0$, namely by the integration constant of
$F_\pi$ and one constant from $\zeta\bdminus$ and $\lambda_0\bminus$. Both of them appear in eqs.\
\eqref{eq:sol13} and \eqref{eq:sol14} but it was seen there that
physics do not depend on the value of these constants, but they can be
absorbed by simple redefinitions of other free variables. In the present
case the situation is different as $\lambda_0\bminus$ determines the
value of $\hat{\chi}\bminus$ and e.g.\ labels states
with different soul contributions to the charge $C_\pi$. To reduce the configuration space of
the solution at $C=0$ to the one of $C\neq0$ in the present setup
one has to choose
$\lambda_0\bminus=0$. Comparison with \eqref{eq:sol7.1} for $C=0$
shows that this condition is exactly \eqref{eq:soln11}. The remaining
constant from $F_\pi$ automatically disappears once this constraint is
imposed. We have argued above that $\lambda_0\bminus$
replaces the anti-commuting Casimir function that was found in
$N=(1,1)$ supergravity at $C=0$. Remarkably enough we now find, that
the reduction of the configuration space implies that this constant
of motion vanishes.

In summary the solutions for $C=0$ can be divided
into two classes: The first class consists of solutions that exist in
an entire neighborhood of $C=0$ and consequently the configuration
space has the same dimension as the one for $C\neq0$. 
However, there
exist additional solutions that appear due to the integration
constants of $F_\pi$ and $\lambda_0\bminus$. These solutions cannot be
extended to the case $C\neq0$ with non-vanishing body.

\subsection{Pure soul Casimir}
\label{sec:sol.3}

This case lies in-between the cases \ref{sec:sol.1} and
\ref{sec:sol.2}. As eq.\ \eqref{eq:sol7.1} for non-invertible $C\neq0$ cannot be solved for
$\zeta\bdminus$ the gauge potential $\tilde{A}\bdminus$ cannot be
expressed in terms of the target-space variables. Therefore, the
solution is parametrized by the same quantities as in the case
$C=0$. The discussion of the two classes of solutions still applies
and again the class of solutions with $\lambda_0\bminus = 0$ can be
obtained smoothly from solutions with non-vanishing body of $C$. Nevertheless
it is important to notice that this does no longer imply the
constraint \eqref{eq:soln11}, as $\hat{\chi}\bminus$ is at least
partially defined through $\bar{\zeta}\bdminus$. 

\section{Solution for general models}
\label{sec:gensol}
Our main task is to solve $N=(2,2)$ supergravity with $Z\neq0$, i.e.\ the models described in
Section \ref{sec:gPSM}. Their solutions can be obtained by applying conformal
transformations interpreted as target space diffeomorphisms to the
solutions at $Z=0$ of the previous Section. In the present Section the variables of the
general model of Section \ref{sec:gPSM} now are underlined (cf.\
  footnote \ref{foot:1}). According to the formulas of Section 4 in
\cite{Bergamin:2004sr} with $Z=Q'$ the new gauge potentials become
\begin{gather}
\label{eq:pcs4}
\begin{align}
  \underline{\omega} &= \omega + \inv{4} \bigl((Z + \bar{Z}) X^b e_b + \bar{Z} \chi
  \psi + Z \bar{\chi} \bar{\psi}  \bigr) \ , & \underline{B} &= B - \frac{i}{4} (\bar{Z} \chi
  \psi - Z \bar{\chi} \bar{\psi}) \ ,\end{align}\medsp
\label{eq:pcs5}
\begin{align}
  \underline{e}_a &= e^{(Q + \bar{Q})/4} e_a \ , & \underline{\psi}_\alpha &= e^{\bar{Q}/4}
  \psi_\alpha \ , & \underline{\Bar{\psi}}_\alpha &= e^{Q/4}
  \bar{\psi}_\alpha\ ,
\end{align}
\end{gather}
with the conformal factor $Q$ being an analytic function in $X = \phi
+ i \pi$. The general solution is obtained by taking these linear
combinations of the solution of the simplified model in eqs.\
\eqref{eq:sol10.1}-\eqref{eq:sol10.8}. At the same time the target
space variables that parametrize these solutions must be transformed
according to
\begin{align}
  \label{eq:pcs3}
  \underline{X} &= X\ , & \underline{X}^a &= e^{-(Q + \bar{Q})/4} X^a\ , & \underline{\chi}^\alpha &=
  e^{-\bar{Q}/4} \chi^\alpha\ , & \underline{\Bar{\chi}}^\alpha &=   e^{-Q/4}
  \bar{\chi}^\alpha\ .
\end{align}
The prepotential transforms as $\underline{u} = e^{\bar{Q}/2} u$.
The definition of the free functions $\extd F$, $F_\pi$,
$\lambda_0\bminus$ and $\zeta\bdminus$ remain unchanged, but the relations \eqref{eq:sol5}-\eqref{eq:sol7.1} must be adjusted
due to eq.\ \eqref{eq:pcs3}. The constant of motion in \eqref{eq:sol8}
changes in such a way that it coincides with \eqref{eq:pcs9.2} in the
case of $C\neq0$ with non-vanishing body. The latter Casimir function
is given by \eqref{eq:pcs9.1}.

The main characteristics of the three classes of solutions as
discussed in Sections \ref{sec:sol.1}, \ref{sec:sol.2} and
\ref{sec:sol.3} remain unchanged. For $C\neq0$ with non-vanishing body
eq.\ \eqref{eq:sol7.1} can be solved for
$\underline{\hat{\chi}}\bminus$. For $C=0$ now
\begin{equation}
\label{eq:pcs6}
\lambda_0\bminus = e^{-\half{i} F_\pi + \inv{4}\bar{Q}} \bigl( e^{\inv{4}(Q+\bar{Q})}
\underline{\tilde{\chi}}\bminus  -
\frac{i \sigma \bar{u}}{2 \sqrt{2}} \underline{\Tilde{\bar{\chi}}}\bplus \bigr)
\end{equation}
is the anti-commuting constant of motion.

Finally it should be mentioned that all the models considered so far were
related to chiral gauging. Twisted chiral gaugings are obtained \cite{Bergamin:2004sr} by the change
of variables  $\chi^- \leftrightarrow
\bar{\chi}^-$ and $\psi_- \leftrightarrow \bar{\psi}_-$. As this
redefinition is defined globally, the discussion of the twisted chiral
case is completely analogous to the one of chiral gauging.

\section{Ungauged supergravity}
\label{sec:ungauged}
Beside the two versions of minimally gauged $N=(2,2)$ supergravity discussed
so far ungauged versions have been found in the context of superstring
compactifications
\cite{Gates:2000fj,Haack:2000di,Berkovits:2001tg}. It was shown by us in
\cite{Bergamin:2004sr} that such models can be obtained from the
Poisson tensor \eqref{eq:psm10}-\eqref{eq:pcs8.5} by decoupling the
scalar field $\pi$ and its associated gauge field $B$. This is done by
a change of variables; instead of $\pi$ the Casimir function $C_\pi$
is used as a new target space coordinate. Then, as $\{ C_\pi, X^I\} \equiv0$ for
all fields $X^I$, the corresponding part of the Poisson tensor can be dropped and
$u(X,\bar{X})$ and $Q(X)$ become functions of the dilaton and the dilatinos\footnote{In a
  more mathematical language, a fixed symplectic leaf with respect to
the foliation by $C_\pi$ is chosen. Thus the ungauged model has a smaller
configuration space than the gauged model, which includes all
symplectic leaves.}.

Again we restrict the explicit calculations to dilaton
prepotential supergravity $Z=0$. As $\pi$ appears in the prepotential $u(\phi+i \pi)$ and $\bar{u}(\phi-i\pi)$ the relevant replacement is
\begin{align}
\label{eq:ug1.1}
  \begin{split}
    u(\phi+i\pi) &= \uhat + \inv{4C_B} \uhat' \bigl(\ubhat \chi^2 - \uhat \bar{\chi}^2 + 4i X^a (\chi \gamma_a \gthree \bar{\chi})\bigr) \medsp
    &\quad + \inv{16 C_B} \chi^2 \bar{\chi}^2 \bigl(\uhat'' + \inv{C_B} (\uhat \ubhat' - \uhat' \ubhat)\bigr)\ .
  \end{split}
\end{align}
Here $\hat{u}(\phi + i C_\pi)$ is the prepotential after the
replacement $\pi \rightarrow C_\pi$ and $C_B = 8 Y -
\hat{u}\Hat{\bar{u}}$ is the body of the Casimir function $C$ with
respect to the ungauged model. 

To determine the solution of the ungauged model we could start from the explicit expansion of the prepotential in terms of the Casimir $C_\pi$ in \eqref{eq:ug1.1}. Then we could determine the solution in terms of the new coordinates
\begin{equation}
\label{eq:sol16}
\check{X}^I = (C, C_\pi, \phi, \lambda, \tilchi\bplus, \tilbchi\bplus, \hatchi\bminus, \hatbchi\bminus)\ .
\end{equation}
This should reproduce the solution of Section \ref{sec:sol.1} and, after
dropping the Casimir $C_\pi$, lead to the solution for the ungauged model as
well. However, the calculation of the corresponding brackets is very
complicated as one has to expand the prepotential in \eqref{eq:sol1.2} as well.

Fortunately, there exists a simple trick to obtain the solution in a straightforward way. We can view the replacement $\pi \rightarrow C_\pi$ as a target
space diffeomorphism and simply apply the ensuing transformation rules of the
gauge fields to the solution obtained in Section \ref{sec:solution.2}. From
the expansion of $C_\pi$ in terms of the variables $\tilde{X}^I$ (cf.\ the
paragraph below eq.\ \eqref{eq:sol10.1})
\begin{equation}
  \label{eq:sol17}
  C_\pi = \pi - \frac{\sigma}{4 \sqrt{2}} \tilchi\bplus \tilbchi\bplus - \frac{\sqrt{2} \sigma}{C} \hatchi\bminus \hatbchi\bminus
\end{equation}
and the solution of Section \ref{sec:sol.1} one finds
\begin{gather}
  \label{eq:sol18.1}
\begin{align}
  \check{A}_{C_\pi} &= \tilde{A}_\pi = - \extd F_\pi\ , &
  \check{A}_\lambda &= \extd \phi\ , & \check{A}_\phi &= - \extd \lambda\ ,
\end{align}\medsp
  \label{eq:sol18.2}
  \check{A}_C = \tilde{A}_C + \frac{\sqrt{2} \sigma}{C^2}
  \hatchi\bminus \hatbchi\bminus \extd F_\pi = - \extd F +
  \frac{\sqrt{2} i \sigma}{C^2} (\hatbchi\bminus \extd \hatchi\bminus +
  \hatchi\bminus \extd \hatbchi \bminus)\ , \medsp
  \label{eq:sol18.3}
  \check{A}\bdplus = \tilde{A}\bdplus - \frac{\sigma}{4 \sqrt{2}}
  \tilbchi\bplus \extd F_\pi = \frac{i \sigma}{2 \sqrt{2}} \extd
  \tilbchi\bplus\ , \medsp
    \label{eq:sol18.4}
  \check{\Bar{A}}\bdplus = \Tilde{\Bar{A}}\bdplus + \frac{\sigma}{4
  \sqrt{2}} \tilchi\bplus \extd F_\pi = \frac{i \sigma}{2 \sqrt{2}}
  \extd \tilchi\bplus\ , \medsp
    \label{eq:sol18.5}
  \check{A}\bdminus = \tilde{A}\bdminus - \frac{\sqrt{2} \sigma}{C}
  \hatbchi\bminus \extd F_\pi = i \frac{2 \sqrt{2} i \sigma}{C} \extd
  \hatbchi\bminus\ , \medsp
    \label{eq:sol18.6}
  \check{\bar{A}}\bdminus = \Tilde{\Bar{A}}\bdminus + \frac{\sqrt{2}
  \sigma}{C} \hatchi\bminus \extd F_\pi = - \frac{2 \sqrt{2} i \sigma}{C}
  \extd \hatchi\bminus\ .
\end{gather}
As expected $\extd F_\pi$ does no longer appear in the transformed expressions---except in the first equation of \eqref{eq:sol18.1} of course---and thus $C_\pi$ may be eliminated consistently.

To transform this solution to the defining coordinates of the ungauged model the prepotential in \eqref{eq:sol1.2} must be replaced by the expansion \eqref{eq:ug1.1}:
\begin{align}
  \label{eq:sol19.1}
  \hatchi\bminus &= \tilchi\bminus - \frac{i \sigma \uhat}{2 \sqrt{2}}
  \tilbchi\bplus - \frac{i \sigma}{4 \sqrt{2} C_B} \uhat' \ubhat
  \tilchi\bminus \tilchi\bplus \tilbchi\bplus + \inv{2 C_B} \uhat'
  \tilchi\bminus \tilbchi\bminus \tilbchi\bplus
\end{align}
Now the original gauge fields of the ungauged models can be obtained by the transformation rules of the target space diffeomorphisms. Notice that $C_B$ is a function of $Y$ and $\phi$, derivatives must be taken with respect to these variables as well. Also, the Lorentz invariant combinations for the spinors must be expressed again in terms of the original fields.

It was pointed out already in ref.\ \cite{Bergamin:2004sr} that the
ungauged model allows for a restricted class of solutions with $C=0$
only. This can be made more explicit at this point: The solutions of
the ungauged model correspond to those of the gauged one where $C_\pi$ can be expressed
in terms of target-space coordinates alone (cf.\ the discussion in Section
\ref{sec:sol.2}). But it was found in the previous Section that these
are exactly the solutions with $\lambda_0\bminus = 0$. Therefore for
the ungauged model the
configuration space at $C=0$ is the same as for $C\neq0$, the
additional solutions found in the minimally gauged model
disappear. The ensuing restriction can be made manifest from eq.\
\eqref{eq:sol19.1}. For $C=0$ one obtains the condition
$\hat{\chi}\bminus=0$, for $C$ pure soul this variable is related to
the Casimir by \eqref{eq:sol7.1}. Though these relations have the same
origin as in the minimally gauged model it should be realized that the
solution in terms of the physical coordinates are different, as the
prepotential must be expanded in terms of $C_\pi$ in the present case.

\section{BPS solutions}
\label{sec:bps}
Solutions that preserve some of the supersymmetries play an important r\^{o}le
in many different aspects of supergravity theories. In \cite{Bergamin:2003mh}
it was shown for $N=(1,1)$ that the gPSM approach to dilaton supergravity is very powerful
in the discussion of BPS states as well. In this Section we present first
steps of an extension to minimally gauged $N=(2,2)$ supergravity. Beside
technical complications the main difference is the appearance of new bosonic
fields and of an additional bosonic Casimir function.

In contrast to \cite{Bergamin:2003mh} our present discussion is restricted to bosonic
field configurations only. Such a configuration is BPS if the supersymmetry
variations of the fermionic variables vanish. From \eqref{eq:symtrans} these
transformations in this simplified case are:
\begin{gather}
\label{eq:tr1}
  \delta \chi^+ = - 2 \sqrt{2} X^{++} \bar{\ve}_+ - u \ve_-\dbsp
\label{eq:tr2}
  \delta \chi^- = - 2 \sqrt{2} X^{--} \bar{\ve}_- - u \ve_+
 \dbsp
\label{eq:tr3}
  \delta \psi_+ = - D \ve_+ + \sqrt{2} e^{- Q/2} \bar{w}' \bar{\ve}_- e_{++} +
  \half{\bar{Z}} X^{--} e_{--} \ve_+ \dbsp
\label{eq:tr4}
  \delta \psi_- = - D \ve_- - \sqrt{2} e^{- Q/2} \bar{w}' \bar{\ve}_+ e_{--} -
  \half{\bar{Z}} X^{++} e_{++} \ve_-
\end{gather}
\subsection{Full supersymmetry}
States that respect all supersymmetries must have
$X^a = 0$. Furthermore the complex dilaton $X=\phi + i \pi$ must be chosen such that
$u(\itindex{X}{BPS},\itindex{\bar{X}}{BPS}) = 0$ and
$u'(\itindex{X}{BPS},\itindex{\bar{X}}{BPS}) = 0$. Solutions of this type are
invariant under all supersymmetries if the transformations parameters are
covariantly constant. The Casimir function $C$ in \eqref{eq:pcs9.1} vanishes
for this solution. The requirement that the fully supersymmetric state is a
ground state fixes the additive ambiguity in the definition
\eqref{eq:pcs9.1}. The Casimir related to the $\uone$ is not restricted to a
specific value. Its possible values depend on the details of the prepotential
and are determined by the condition $u = u' = 0$.

The eom of the complex dilaton reduces to $\extd X = 0$ and therefore the solutions
belong to the special class of ``constant dilaton vacua'' CDV
\cite{Grumiller:2003ad,Bergamin:2003mh}. As $u'=0$ the eom of the spin
connection reduces to $\extd \omega = 0$ and thus curvature vanishes. 
\subsection{BPS states}
In many applications the interesting field configurations are restricted to
vanishing fermion fields. Eqs.\ \eqref{eq:tr1} and \eqref{eq:tr2} imply
\begin{align}
  \label{eq:bps1}
  u \ve_- &= - 2 \sqrt{2} X^{++} \bar{\ve}_+\ , &  u \ve_+ &= - 2 \sqrt{2} X^{--}
  \bar{\ve}_-\ .
\end{align}
Iteration of these equations (and their hermitian conjugates) imply that
$C=0$. This is equivalent to the statement that the determinant of the purely
fermionic part of the Poisson tensor must vanish. There exist three different
types of solutions.
\subsubsection{CDV solutions}
\label{sec:cdv}
If $X^{++}=X^{--}=0$ the complex dilaton $X$ is again constant and the
prepotential vanishes on the solution. However $u' \neq 0$, else the fully
supersymmetric solution would be recovered. from the eom for the spin
connection one deduces
\begin{equation}
  \label{eq:bps2}
  R = 2 \ast \extd \omega = \half{1} u' \bar{u}' > 0\ ,
\end{equation}
which in our conventions implies AdS space. The solutions do not respect full
supersymmetry as from \eqref{eq:tr3} and \eqref{eq:tr4}
\begin{align}
  \label{eq:bps3}
  D \ve_+ - \inv{2 \sqrt{2}} \bar{u}' e_{++} \bar{\ve}_- &= 0\ , & D \ve_- +
  \inv{2 \sqrt{2}} \bar{u}' e_{--} \bar{\ve}_+ &= 0\ .
\end{align}
\subsubsection{Chiral solutions}
\label{sec:mink}
Even with the choice $X^{++} \neq 0$ it may happen that $X^{--} = u = 0$. In
that case $\ve_+ = 0$ and only the $\ve_-$ component can be non-zero. As for all
solutions with vanishing fermions $\extd \pi = 0$, but now $\extd \phi \neq
0$. Together with the condition $u=0$ on the solution this however implies $u
\equiv 0$. Therefore all chiral solutions are flat Minkowski space. As $X^{++}
\neq 0$ this case is covered by the discussion of Section
\ref{sec:gensol}. Integration of eq.\ \eqref{eq:tr4} yields
\begin{equation}
  \label{eq:bps4}
  \ve_- = \exp\bigl[\half{1} (\bar{Q} - i F_\pi) \bigr] \sqrt{|X^{++}|} \tilde{\ve}
\end{equation}
with a constant spinor $\tilde{\ve}$. All states of this type respect two of
the four supersymmetries.
\subsubsection{Supersymmtric black holes}
Obviously the cases \ref{sec:cdv} and \ref{sec:mink} do not describe
supersymmetric black hole solutions. The latter only exist if all three
quantities $X^{++}$, $X^{--}$ and $u$ are different from zero. Then from
\eqref{eq:bps1} it follows that both components $\ve_+$ and $\ve_-$ must be
nonvanishing. It remains to check the differential equations
\eqref{eq:tr3} and \eqref{eq:tr4}. By the use of the explicit solution derived in
the previous Sections it can be shown that \eqref{eq:tr4} is the hermitian
conjugate of \eqref{eq:tr3}. Thus it remains to solve a single differential equation that can be
written as
\begin{equation}
  \label{eq:bps5}
  \extd \ve_- = \extd \bigl(\half{1} \ln X^{++} - \half{i} F_\pi + \half{1}
  \bar{Q} \bigr) \ve_- + 2 (Z-\bar{Z}) W \extd F \ve_-\ .
\end{equation}
If $Z$ is real the solution for $\ve_-$ is given by eq.\ \eqref{eq:bps4} while
$\ve_+$ reads:
\begin{equation}
  \label{eq:bps6}
  \ve_+ = - \frac{i \sigma}{2 \sqrt{2}} \exp\bigl[\half{1}(\bar{Q} - i F_\pi)  \bigr]
  \frac{u}{\sqrt{|X^{++}|}} \tilde{\ve}
\end{equation}
In the general case a closed analytic expression cannot be obtained. Again all
solutions respect half of the supersymmetries.

As a result of this Section it
follows that any bosonic field configuration with $C=0$ \emph{locally} is BPS. At the same time it should be realized that this need not be
true globally. Indeed, global solutions in the general case are obtained by
a combination of several patches, which may destroy the BPS property at
the global level (cf.\ \cite{Bergamin:2004me}). Finally we mention the
agreement of these calculations with general statements on supersymmetric
black hole solutions \cite{Gibbons:1981ja,Gibbons:1982fy,Tod:1983pm}: all
supersymmetric black holes are extremal. In our calculations this immediately
follows from the Killing norm for $C=0$ \cite{Grumiller:2002nm,Bergamin:2003mh}
\begin{equation}
  \label{eq:bps7}
  K(X) = -2 e^{(Q+\bar{Q})/2} W = \inv{4} \bigl| e^{\bar{Q}} u \bigr|^2\ .
\end{equation}
Obviously all zeros are of even degree.

\section{Conclusions}
\label{sec:conclusions}
The present work on $N=\left(2,2\right)$ supergravity in two dimensions
extends our previous one \cite{Bergamin:2004sr} by providing for the first time the
full classical solutions, including the complete fermionic parts.
This is possible thanks to the powerful tool of the equivalent formulation
as a particular class of graded Poisson Sigma models. Although the
actual computation is restricted to the chiral case, the twisted chiral
$N=\left(2,2\right)$ theories, as well as the ungauged version can
be obtained by simple redefinitions in the formulas presented here.
The classification of the solutions is determined by the values of
the Casimir functions in which the interplay between body and soul characterizes different cases.
Although we draw heavily from our experience with the $N=\left(1,1\right)$
case \cite{Ertl:2000si,Bergamin:2003am}, the present results exhibit new
interesting structures due to the larger fermionic symmetry algebra.

As yet another application of research directions which are possible
in the framework of gPSMs we discuss solutions retaining certain
supersymmetries (BPS states), although in $N=\left(2,2\right)$ supergravity
the analysis turns out to be rather more involved than the one in
$N=\left(1,1\right)$. Therefore, only bosonic solutions are treated
here.

Comparing with other results, already obtained for $N=\left(1,1\right)$
it is clear that beside a more comprehensive discussion of BPS states
in analogy to ref.\ \cite{Bergamin:2003mh}, also the problem of putting a (super)
point particle into an $N=\left(2,2\right)$ background (cf.\ \cite{Bergamin:2003am}
for $N=\left(1,1\right)$), the coupling of supersymmetric matter
(cf.\ \cite{Bergamin:2003mh}) as well as the quantization of $N=\left(2,2\right)$
supergravity (cf.\ \cite{Bergamin:2004us,Bergamin:2004aw}) are topics expected to allow a successful treatment in
further work. This will allow new insights also for application in
superstring theory where the gPSM approach now seems to provide a new
line of attack for the solution of some old problems.

\subsection*{Acknowledgement}
The authors would like to thank  D.~Grumiller, P.~van~Nieuwenhuizen,
E.~Scheidegger, T.~Strobl and D.~Vassilevich for fruitful discussions. L.B.\ acknowledges the hospitality of the Max-Planck-Institut f\"{u}r
Gravitationsphysik, where a part of this work has been developed.
This work has been supported by the project P-16030-N08 of the
Austrian Science Foundation (FWF).

\appendix
\section{Notations and conventions}
\label{sec:notation}
The conventions are identical to
\cite{Ertl:2000si,Ertl:2001sj}, where additional explanations can be found.

Indices chosen from the Latin alphabet are generic (upper case) or (lower
case) refer to commuting objects, Greek indices are anti-commuting ones. Holonomic coordinates
are labeled by $M$, $N$, $O$ etc., anholonomic ones by $A$, $B$, $C$ etc., whereas
$I$, $J$, $K$ etc.\ are general indices of the gPSM:
  \begin{align}
    X^I &= (X^\phi, X^\pi, X^a, X^\alpha, X^{\bar{\alpha}}) = (\phi, \pi, X^a,
    \chi^\alpha , \bar{\chi}^\alpha)\medsp
    A_I &= (A_\phi, A_\pi, A_a, A_\alpha, A_{\bar{\alpha}}) = (\omega, B, e_a,
    \psi_\alpha, \bar{\psi}_\alpha)
  \end{align}

The summation convention is always $NW \rightarrow SE$, e.g.\ for a
fermion $\chi$: $\chi^2 = \chi^\alpha \chi_\alpha$. Our conventions are
arranged in such a way that almost every bosonic expression is transformed
trivially to the graded case when using this summation convention and
replacing commuting indices by general ones. This is possible together with
exterior derivatives acting \emph{from the right}, only. Thus the graded
Leibniz rule is given by
\begin{equation}
  \label{eq:leibniz}
  \mbox{d}\left( AB\right) =A\mbox{d}B+\left( -1\right) ^{B}(\mbox{d}A) B\ .
\end{equation}

In terms of anholonomic indices the metric and the symplectic $2 \times 2$
tensor are defined as
\begin{align}
  \eta_{ab} &= \left( \begin{array}{cc} 1 & 0 \\ 0 & -1
  \end{array} \right)\ , &
  \epsilon_{ab} &= - \epsilon^{ab} = \left( \begin{array}{cc} 0 & 1 \\ -1 & 0
  \end{array} \right)\ , & \epsilon_{\alpha \beta} &= \epsilon^{\alpha \beta} = \left( \begin{array}{cc} 0 & 1 \\ -1 & 0
  \end{array} \right)\ .
\end{align}
The metric in terms of holonomic indices is obtained by $g_{mn} = e_n^b e_m^a
\eta_{ab}$ and for the determinant the standard expression $e = \det e_m^a =
\sqrt{- \det g_{mn}}$ is used. The volume form reads $\epsilon = \half{1}
\epsilon^{ab} e_b \wedge e_a$; by definition $\ast \epsilon = 1$.

The $\gamma$-matrices are used in a chiral representation:
\begin{align}
\label{eq:gammadef}
  {{\gamma^0}_\alpha}^\beta &= \left( \begin{array}{cc} 0 & 1 \\ 1 & 0
  \end{array} \right) & {{\gamma^1}_\alpha}^\beta &= \left( \begin{array}{cc} 0 & 1 \\ -1 & 0
  \end{array} \right) & {{\gthree}_\alpha}^\beta &= {(\gamma^1
    \gamma^0)_\alpha}^\beta = \left( \begin{array}{cc} 1 & 0 \\ 0 & -1
  \end{array} \right)
\end{align}

Covariant derivatives of anholonomic indices with respect to the geometric
variables $e_a = \extd x^m e_{am}$ and $\psi_\alpha = \extd x^m \psi_{\alpha m}$
include the two-dimensional spin-connection one form $\omega^{ab} = \omega
\epsilon^{ab}$. When acting on lower indices the explicit expressions read
($\half{1} \gthree$ is the generator of Lorentz transformations in spinor space):
\begin{align}
\label{eq:A8}
  (D e)_a &= \extd e_a + \omega {\epsilon_a}^b e_b & (D \psi)_\alpha &= \extd
  \psi_\alpha - \half{1} {{\omega \gthree}_\alpha}^\beta \psi_\beta
\end{align}

Dirac conjugation is defined as $\bar{\chi}^\alpha = \chi\dega
\gamma_0$. Written in components of the chiral representation 
\begin{align}
\label{eq:Achi}
  \chi^\alpha &= ( \chi^+, \chi^-)\ , & \chi_\alpha &= \begin{pmatrix} \chi_+ \\
  \chi_- \end{pmatrix}
\end{align}
the relation between upper and lower indices becomes $\chi^+ = \chi_-$,
$\chi^- = - \chi_+$. Dirac conjugation follows as $\bar{\chi}_- =
\chi_-\cc$, $\bar{\chi}_+ = - \chi_+\cc$, i.e.\ for Majorana spinors $\chi_-$ is real while $\chi_+$ is imaginary.

For two gauge-covariant Dirac spinors $\chi_\alpha$ and $\lambda_\alpha$ the
combinations
\begin{alignat}{3}
\chi \lambda\ , &\qquad&\qquad \chi \gthree \lambda\ , &\qquad&\qquad \bar{\chi} \gamma^a \lambda
\end{alignat}
and their hermitian conjugates are gauge invariant for chiral gaugings, while
\begin{alignat}{3}
\label{eq:A11}
\bar{\chi} \lambda\ , &\qquad&\qquad \bar{\chi} \gthree \lambda\ , &\qquad&\qquad \bar{\chi} \gamma^a \lambda
\end{alignat}
are invariant for twisted-chiral gaugings. Note that in the latter case the
gravitino $\psi_\alpha$ transforms under gauge transformations as
$\bar{\chi}_\alpha$. Thus in eq.\ \eqref{eq:A11} the bilinear invariants of a gravitino and a
dilatino are obtained by substituting $\lambda \rightarrow \bar{\psi}$.

Vectors in light-cone coordinates are given by
\begin{align}
\label{eq:A10}
  v^{++} &= \frac{i}{\sqrt{2}} (v^0 + v^1)\ , & v^{--} &= \frac{-i}{\sqrt{2}}
  (v^0 - v^1)\ .
\end{align}
The additional factor $i$ in \eqref{eq:A10} permits a direct identification of the light-cone components with
the components of the spin-tensor $v^{\alpha \beta} = \frac{i}{\sqrt{2}} v^c \gamma_c^{\alpha
  \beta}$. This implies that $\eta_{++|--} = 1$
and $\epsilon_{--|++} = - \epsilon_{++|--} = 1$. The
$\gamma$-matrices in light-cone coordinates become
\begin{align}
\label{eq:gammalc}
  {(\gamma^{++})_\alpha}^\beta &= \sqrt{2} i \left( \begin{array}{cc} 0 & 1 \\ 0 & 0
  \end{array} \right)\ , & {(\gamma^{--})_\alpha}^\beta &= - \sqrt{2} i \left( \begin{array}{cc} 0 & 0 \\ 1 & 0
  \end{array} \right)\ .
\end{align}

\providecommand{\href}[2]{#2}\begingroup\raggedright\endgroup

\end{document}